\documentclass[twocolumn,prl,10pts,nofootinbib,superscriptaddress,preprintnumbers]{revtex4-1}
\usepackage[utf8]{inputenc}

\def\mySections#1{{\bf #1.} } 
\usepackage{hyperref}
\usepackage{braket}  
\usepackage[dvips]{graphicx}
\usepackage{hhline}
\usepackage{epsfig,amsmath,amssymb,verbatim,mathrsfs,array,layout,textcomp,amssymb,latexsym,slashed,graphicx,booktabs,color,mathtools}
\usepackage{hyperref}

\newcommand{\beq}{\begin{equation}}
\newcommand{\eeq}{\end{equation}}
\newcommand{\nn}{\nonumber}
\def\beqa{\begin{eqnarray}}
\def\eeqa{\end{eqnarray}}
\def\bea{\begin{eqnarray}}
\def\eea{\end{eqnarray}}

\newcommand{\bv}{\left(\begin{array}{c}}
\newcommand{\ev}{\end{array}\right)}
\newcommand{\bmtwo}{\left(\begin{array}{cc}}
\newcommand{\bmthree}{\left(\begin{array}{ccc}}
\newcommand{\emn}{\end{array}\right)}
\newcommand{\bmtwoc}{\left\{\begin{array}{cc}}
\newcommand{\bmthreec}{\left\{\begin{array}{ccc}}
\newcommand{\emnc}{\end{array}\right\}}
\newcommand{\ba}{\begin{array}}
\newcommand{\ea}{\end{array}}
\newcommand{\be}{\begin{eqnarray}}
\newcommand{\ee}{\end{eqnarray}}

\newcommand{\GeV}{\text{ GeV}}

\newcommand{\eV}{\text{ eV}}

\definecolor{readableRTD}{rgb}{0.7,0.1,0.2}

\definecolor{readableMG}{rgb}{0.0,0,0.5}

\def\lsim{\mathrel{\rlap{\lower4pt\hbox{\hskip1pt$\sim$}}
     \raise1pt\hbox{$<$}}}         
\def\gsim{\mathrel{\rlap{\lower4pt\hbox{\hskip1pt$\sim$}}
     \raise1pt\hbox{$>$}}}         

\definecolor{bluDT}{cmyk}{1,0.5,0,0.3}

\definecolor{darkblue}{rgb}{0.2,0.2,0.9}
\definecolor{colorRTD}{rgb}{.2,.2,.7}

\usepackage{soul}

\begin{document}

\preprint{CERN-TH-2021-055}

\title{{\LARGE Sliding Naturalness:} \\[0.5em]  A New Solution to the Strong-CP and Electroweak Hierarchy Problems} 

\author{Raffaele Tito D'Agnolo}
\affiliation{Institut de Physique Th\'eorique, Universit\'e Paris Saclay, CEA, F-91191 Gif-sur-Yvette, France}

\author{Daniele Teresi}
\affiliation{CERN, Theoretical Physics Department, 1211 Geneva 23, Switzerland}

\begin{abstract}
We present a novel framework to solve simultaneously the electroweak hierarchy problem and the strong-CP problem. A small but finite Higgs vacuum expectation value and a small $\theta$-angle are selected after the QCD phase transition, without relying on the Peccei-Quinn mechanism or other traditional solutions. We predict a distinctive pattern of correlated signals at hadronic EDM, fuzzy dark matter and axion experiments. 
\end{abstract}
\maketitle

\section{Introduction}
Our whole understanding of physics is based on symmetry. Among its countless successes, one of the best examples is dimensional analysis, which is just a convenient way of enforcing the selection rules of dilatations.

The most spectacular failures of symmetry are the cosmological constant (CC), the Higgs boson mass squared ($m_h^2$) and the QCD $\theta$-angle. All three are orders of magnitude smaller than what we expect from dimensional analysis and the known symmetries of Nature. 

Taken at face value these three parameters appear completely unrelated. The CC is coupled to gravity and determines the maximal observable size of the Universe. $m_h^2$ is coupled to weak interactions, it determines their range and the mass of most known particles. $\theta$ is coupled to strong interactions and determines a series of properties of mesons and baryons, including the neutron electric dipole moment (EDM). 
If we try to understand these parameters using symmetry in the most direct way we are immediately led to separate the three problems. 
After decades of theoretical and experimental efforts, we have not yet found any evidence that the symmetry paradigms that separate these puzzles are realized in Nature. However there is a deep connection between $m_h^2$ and $\theta$, founded on a more indirect use of symmetry. Only one local operator in the Standard Model (SM) has a vacuum expectation value sensitive to $m_h^2$: $G\widetilde G$, the operator coupled to $\theta$.

In this Letter we present a novel framework that simultaneously explains the small observed value of $m_h^2$ and $\theta$, with the CC playing an important role in their selection. We call it {\it Sliding Naturalness}. 
The selection of $m_h^2$ and $\theta$ is dynamical and occurs in the early history of the Universe, right after the QCD phase transition. Other ideas of ``cosmological naturalness" that select a small value for $m_h^2$ in the early history of the Universe include~\cite{
Dvali:2003br,Dvali:2004tma,Graham:2015cka,Arkani-Hamed:2016rle, Arvanitaki:2016xds,Geller:2018xvz,  Cheung:2018xnu, Giudice:2019iwl, Strumia:2020bdy, Csaki:2020zqz, Arkani-Hamed:2020yna,  Giudice:2021viw}. 

We do not make any assumption on the high-energy physics responsible for the observed homogeneity and isotropy of the Universe. In particular the selection of $m_h^2$ and $\theta$ is compatible with any realization of inflation, but also with modern swampland conjectures~\cite{Palti:2019pca}. Importantly we do not require hidden model building or tuning to enforce a large number of $e$-folds or a low scale of inflation. At low energies, the entire model can be described by a simple polynomial potential for two approximately shift-symmetric axion-like particles, one of which can be the dark matter of our Universe. Explaining the observed $m_h^2$ and $\theta$ does not require super-Planckian field excursions.

We predict a very distinctive phenomenology that relates signals in hadronic EDM experiments, fuzzy dark matter and axion searches, defining a smoking-gun pattern that can precisely identify this framework.

\begin{figure*}[!t]
$$\includegraphics[width=0.3\textwidth]{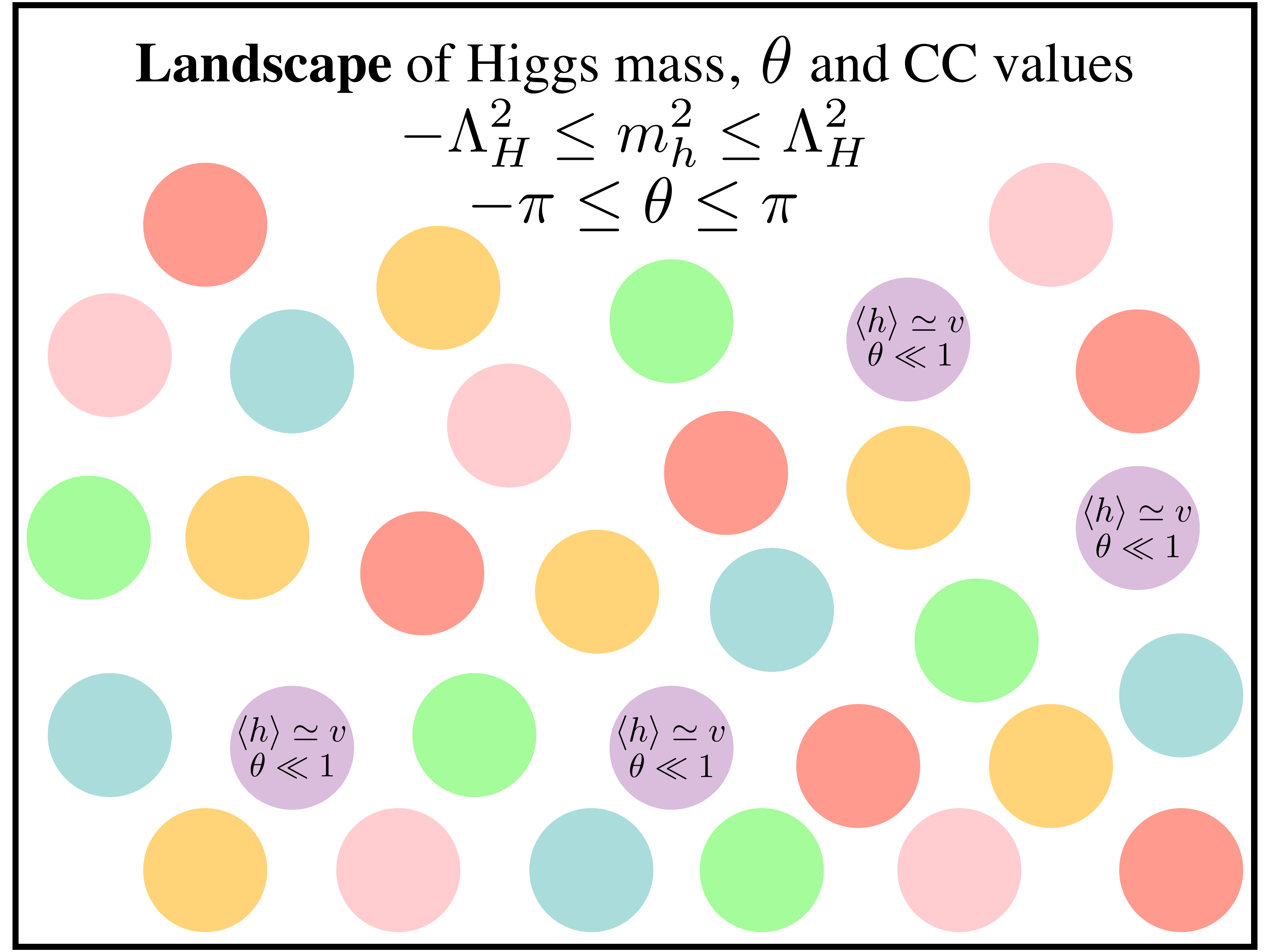} \includegraphics[width=0.3\textwidth]{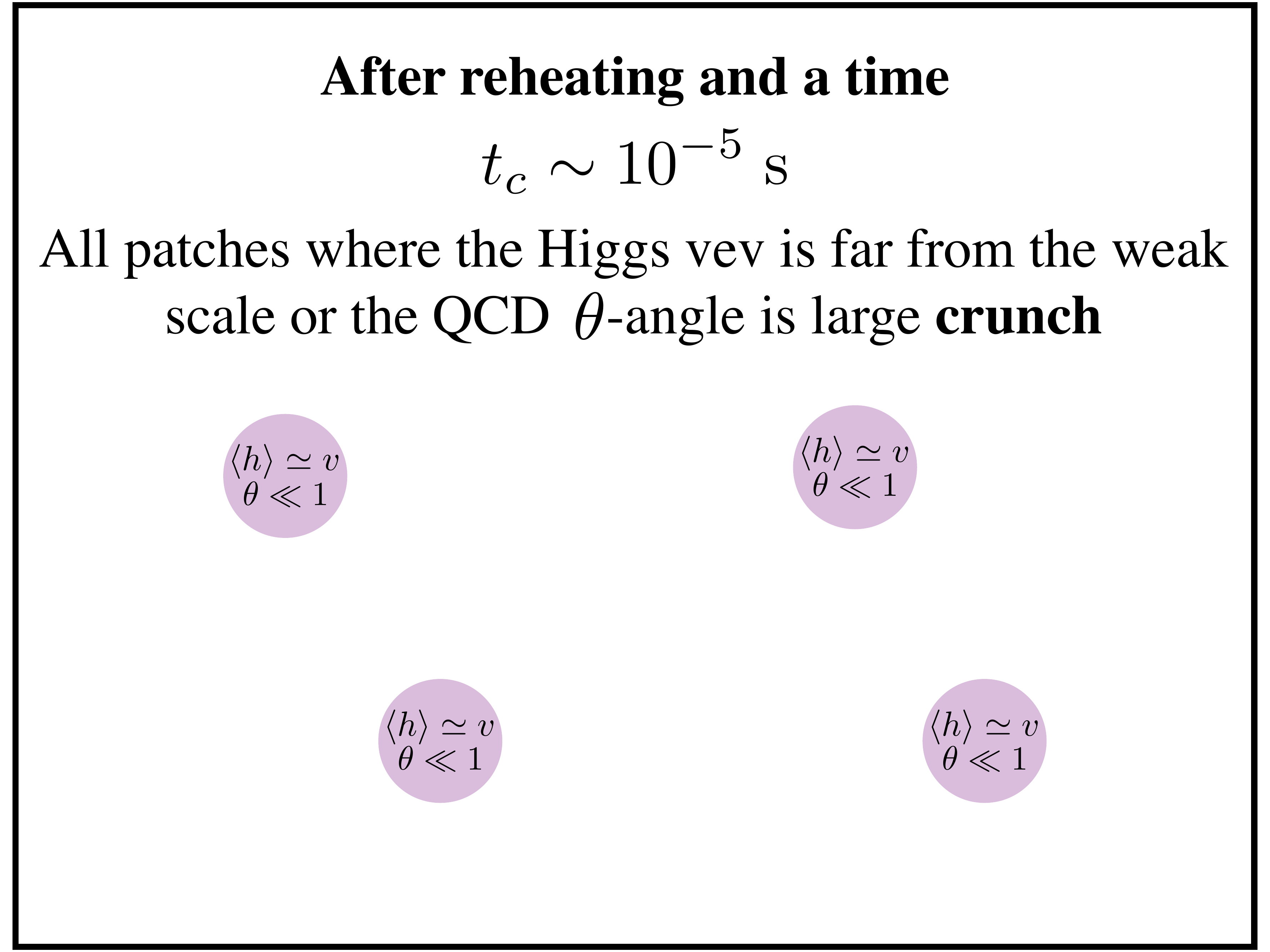} \includegraphics[width=0.3\textwidth]{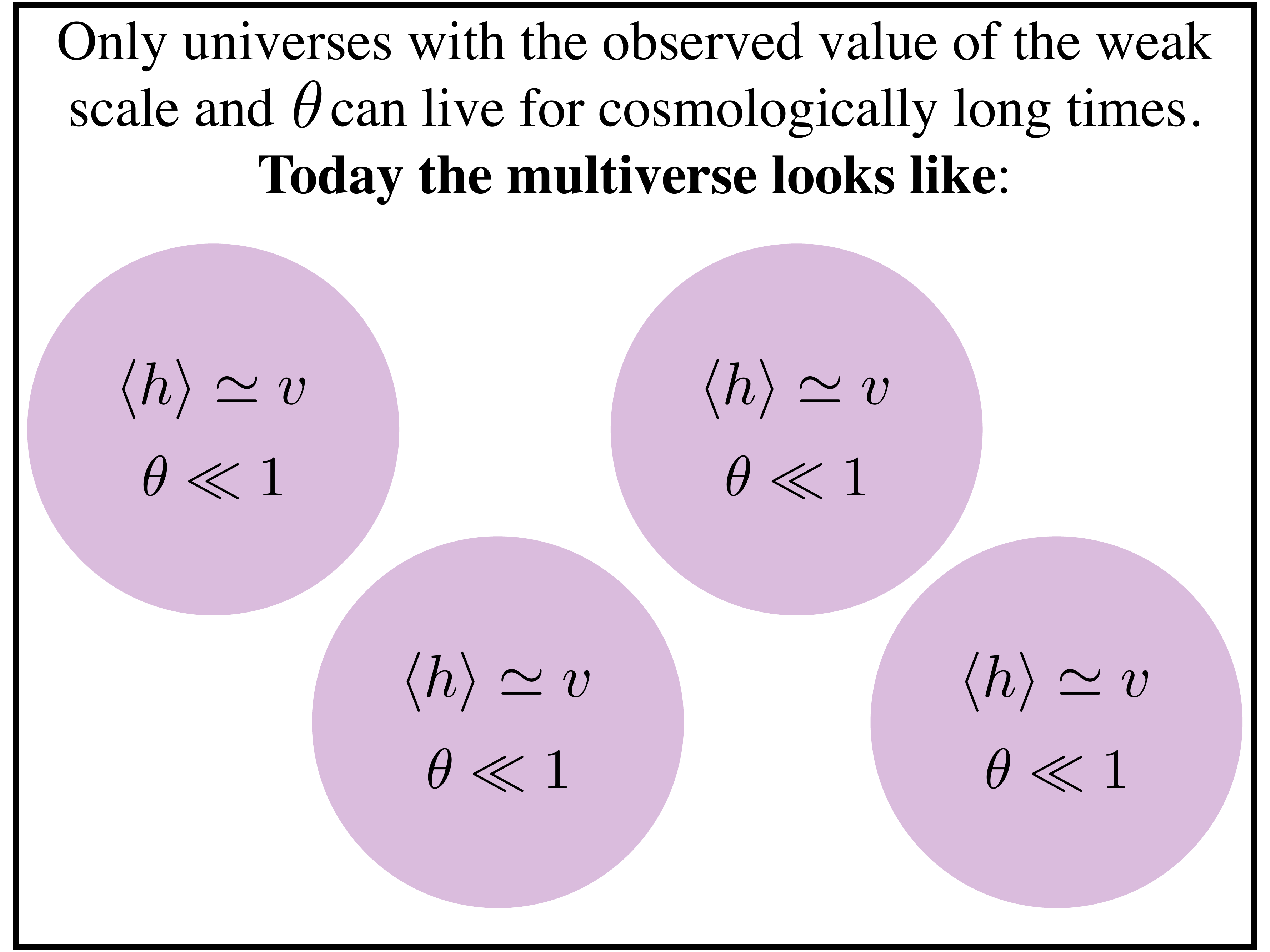} $$
\caption{Cartoon of the basic mechanism: a landscape of Higgs masses and QCD $\theta$-angles is populated early in the history of the Universe. All patches with Higgs vev and $\theta$ far from the observed ones crunch after the QCD phase transition. The only universes that survive until today and grow to cosmological sizes are those with $\langle h \rangle \simeq v$ and $\theta \lesssim 10^{-10}$.\label{fig:mechanism}}
\end{figure*}
\section{Basic Idea}\label{sec:mechanism}

We imagine the existence of a landscape for the Higgs mass squared, the cosmological constant and the QCD $\theta$-angle. Different patches of the Universe have different values for these quantities. We denote with $\Lambda_H^2$ the maximum value of $m_h^2$ in the landscape and with $\Lambda_{\rm max}^4$ the maximum value of the CC. $\Lambda_H$ and $\Lambda_{\rm max}$ can both be $\mathcal{O}(1)$ in units of $M_{\rm Pl}$. 

At low energy the theory includes two new scalars $\phi_\pm$ with an approximate shift symmetry. The $\phi_\pm$ potential has a deep minimum with energy density $\lesssim -\Lambda_{\rm max}^4$. Patches where $\phi_\pm$ roll to their global minimum rapidly crunch~\cite{Strumia:2019kxg}. 

A second metastable minimum for $\phi_\pm$ is generated only if both the Higgs vacuum expectation value (vev) $\langle h \rangle$ and the QCD $\theta$-angle are in specific ranges
\be
\mu_S \leq \langle h \rangle \leq \mu_B\, , \quad \theta \leq \theta_{\rm max}\, , \label{eq:interval}
\ee
where $\mu_B$ can naturally be $\ll \Lambda_H$ and $\theta_{\rm max} \ll 1$. In this minimum the CC can be tuned to zero, for instance because of Weinberg's argument on the existence of galaxies~\cite{Weinberg:1988cp}. This does not require any coincidence in the landscape, as explained in the Appendix. Only universes with \textit{small} and \textit{nonzero} Higgs vev and \textit{small} QCD $\theta$-angle survive until today, as shown schematically in Fig.~\ref{fig:mechanism}. All other universes rapidly crunch. We discuss a similar way to select $m_h^2$ independently of $\theta$ in a companion paper~\cite{DAgnoloTeresi}. Different versions of a crunching mechanism have been advocated to explain the value of the CC in~\cite{Bloch:2019bvc} and of $m_h^2$ in~\cite{Strumia:2020bdy,Csaki:2020zqz}.

\section{Selection of the Weak Scale and of $\theta$}\label{sec:mechanism}
We can realize this idea with a large class of potentials. The simplest choice is $V=V_{\phi}+V_{\phi H}$, with $V_{\phi}=V_{\phi_+} + V_{\phi_-}$ and
\be
V_{\phi_\pm} &=&  \mp \frac{m_{\phi_\pm}^2}{2} \phi_\pm^2 - \frac{m_{\phi_\pm}^2}{4 M_\pm^2}\phi_\pm^4\nn \\
V_{\phi H} &=& -\frac{\alpha_s}{8 \pi}\left(\frac{\phi_+}{F_+}+\frac{\phi_-}{F_-}+\theta\right)\widetilde G G\, , \label{eq:potential}
\ee
where $m_{\phi_\pm}^2>0$ and $G \widetilde G \equiv (\epsilon^{\mu\nu\rho\sigma}/2)G_{\mu\nu}^a G_{\rho\sigma}^a$. $V_{\phi}$ and $V_{\phi H}$ break the shift symmetry $\phi_\pm \to \phi_\pm + c_\pm$ by a small amount $m_{\phi_\pm}, \Lambda_{\rm QCD} \ll F_\pm, M_\pm$. $V_{\phi}$ is an EFT description valid at least for $|\phi_\pm|\lesssim M_\pm$. A concrete way to generate Eq.~\eqref{eq:potential} and UV complete it is described in the Appendix. As discussed there, the strong-CP and hierarchy problems are essentially solved by the approximate symmetries of $\phi_\pm$ that keep the hierarchy between their minima stable. The $\phi_\pm$ sector is very weakly coupled to the SM, where the symmetry is not manifest.

\begin{figure}[t]
$$\hspace{-2em}\includegraphics[width=0.9\columnwidth]{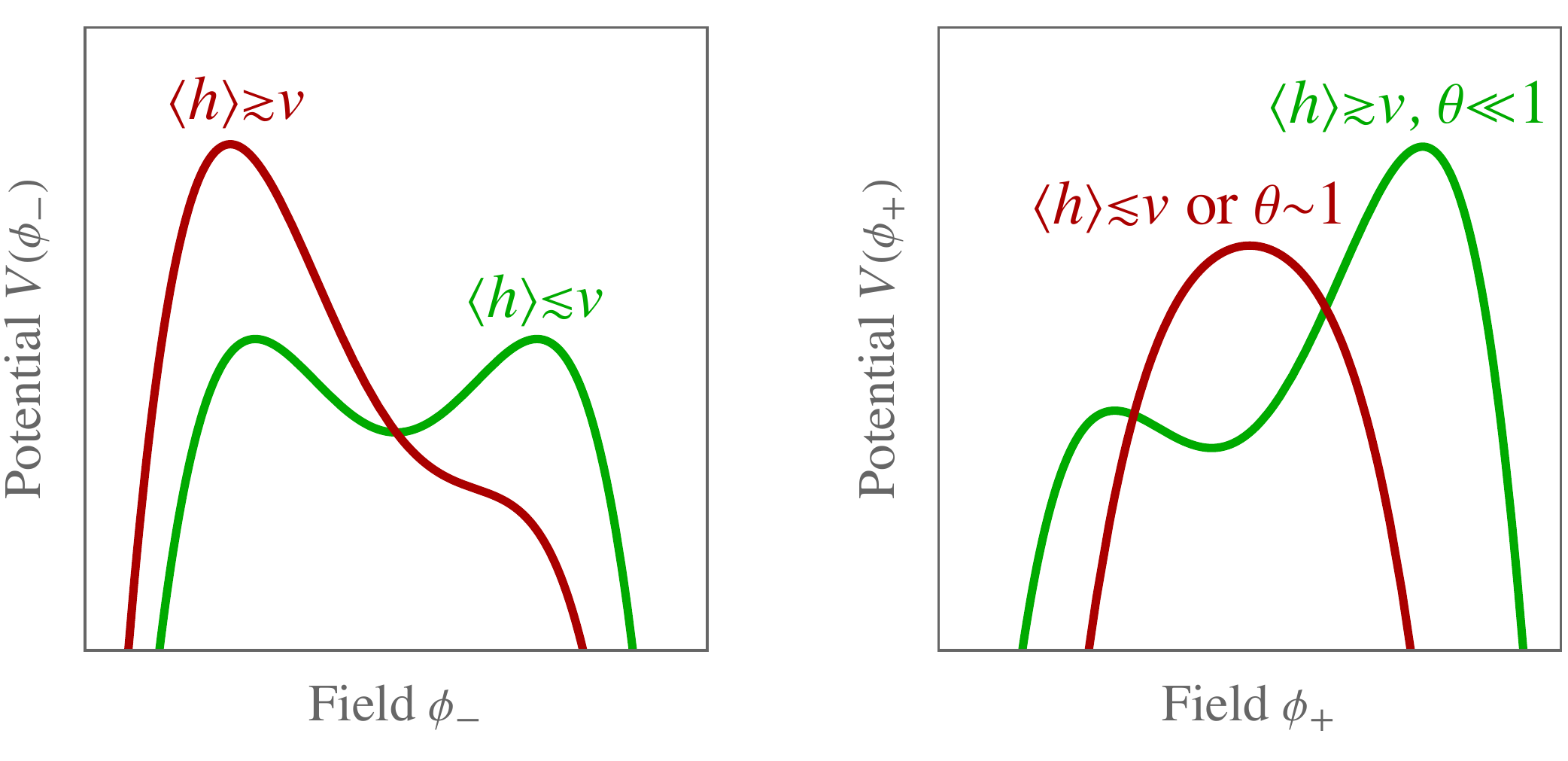}$$
\caption{Example potentials $V(\phi_\pm)$ that select the weak scale. If $\langle h \rangle \gtrsim v$ the local minimum of $V(\phi_-)$ is destabilized (left panel), so that the field rolls down towards its deep global minimum (not shown in the plot) and the Universe rapidly crunches. Conversely, the local minimum of $V(\phi_+)$ is generated only if $\langle h \rangle \gtrsim v$ and $\theta \lesssim \theta_{\rm max}$. \label{fig:potential}}
\end{figure}

$V_{\phi_\pm}$ in Eq.~\eqref{eq:potential} describes two decoupled scalars: $\phi_-$ has a minimum at zero and two maxima at $M_-$, while $\phi_+$ has a maximum at zero and no stable minimum, as shown in Fig~\ref{fig:potential}. It is technically natural to take their cross-couplings to be negligibly small as discussed in the Appendix. Note that one can have the same structure for more general values of the $\mathcal{O}(1)$ factors in Eq.~\eqref{eq:potential}, as well as including tadpole and cubic terms (with scale $M_\pm$) and possibly cross-interactions. We discuss this in~\cite{DAgnoloTeresi}. Notice that, while the potential in Eq.~\eqref{eq:potential} should be considered just as an illustrative example, it is technically natural as it stands.
 
If we include $V_{\phi H}$, a non-zero Higgs vev can generate a safe minimum for $\phi_+$, provided that $\theta$ is sufficiently small. However when the Higgs vev becomes too large $V_{\phi H}$ destabilizes the original safe minimum for $\phi_-$. This is shown in Fig~\ref{fig:potential}. 

To see how this is realized in practice, we can focus on the region $|\phi_\pm| \lesssim M_\pm$ and take $M_\pm/F_\pm \ll 1$. Then if we rotate $\phi_\pm$ in the quark mass matrix and match to the chiral Lagrangian at low energy,
\be
V_{\phi H} \simeq  \frac{\Lambda^4(\langle h\rangle)}{2}\left(\theta + \frac{\phi_-}{F_-}+ \frac{\phi_+}{F_+}\right)^2+\ldots\, . 
\label{eq:phiG}
\ee
For simplicity we have also expanded for $\theta \ll 1$, but our arguments hold also for $\theta=\mathcal{O}(1)$. The above potential is generated at the QCD phase transition and its overall size is a monotonic function of the Higgs vev through chiral symmetry breaking. For $m_{u, d} \lesssim 4\pi f_\pi$ the scale of the potential reads~\cite{Villadoro}
\be
\Lambda^4(\langle h\rangle) = m_\pi^2 f_\pi^2 \frac{m_u m_d}{(m_u+m_d)^2}\, .
\ee
The selection of $\langle h \rangle$ and of $\theta$ takes place after the QCD phase transition, during radiation domination. As discussed in the next Section, it takes $\phi_\pm$ a time $1/m_{\phi_\pm}$ to slide towards the deep minimum and crunch, if the safe local minimum is not present. For simplicity, in the following we take $1/m_{\phi_-} \ll 1/m_{\phi_+}$, so that $\phi_-$ starts rolling first.

The safe local minimum for $\phi_-$ only exists if the QCD tadpole does not destroy it. Given Eq.~\eqref{eq:phi+}, we can take $M_-/F_- \lesssim \theta + M_+/F_+$, so that the tadpole term dominates the $\phi_-$ potential in Eq.~\eqref{eq:phiG} (recall that $\phi_- \sim M_-$). Hence, the cross-interactions between $\phi_\pm$ generated by QCD are negligible and the minimization problem factorizes into two separate ones for the two scalars. As a consequence the local minimum for $\phi_-$ is easily found to exist only if
\be
\Lambda^4(\langle h\rangle) \lesssim \frac{m_{\phi_-}^2M_- F_-}{\left(\theta + M_+/F_+\right)}\, . \label{eq:phi-}
\ee
We have replaced the unknown value of $\phi_+$ at the time when ${\phi_-}$ starts moving, with $M_+$, because all universes where $|\phi_+| \gg M_+$ crunch and in all others this is its typical value. Even in universes where the initial position of $\phi_+$ is tuned to give $\theta+\phi_{+}^{\rm initial}/F_+\simeq 0$, $\phi_-$ still sees an effective $\theta$-angle of $\mathcal{O}(M_+/F_+)$ after $\phi_+$ starts rolling, because $\phi_+$ moves by about $M_+$ after the QCD phase transition\footnote{There is an exception to this argument. In universes that are doubly tuned, i.e. $\theta+\phi_{+}^{\rm initial}/F_+\simeq 0$ and $\phi_{+}^{\rm initial} \simeq \phi_{+,{\rm min}}$ the $\phi_-$ tadpole can be smaller than $\Lambda^4(\langle h\rangle)M_+/(F_-F_+)$, since in this case $\phi_+$ essentially does not move throughout its cosmological history. However, the double tuning makes these patches irrelevant, as discussed in detail in \cite{DAgnoloTeresi}}.  The above inequality is thus an upper bound on $\langle h \rangle$. 

Similarly, a safe local minimum for $\phi_+$ is found to exist only if
\be
\Lambda^4(\langle h\rangle) \gtrsim m_{\phi_+}^2 F_+^2 \quad {\rm and} \quad \frac{M_+}{F_+} \gtrsim \theta + \frac{M_-}{F_-} \, . \label{eq:phi+}
\ee
This puts an upper bound on the maximal neutron EDM that we can observe today and a lower bound on the Higgs vev. The first inequality implies that the positive curvature from $V_{\phi H}$ beats the negative one from $V_{\phi_+}$, while the second one enforces that this happens in the region where the negative quartic from $V_{\phi_+}$ is subleading. Moreover, we have assumed that the first inequality is almost saturated.

We have found that $\phi_+$ gives an upper bound on $\theta$ and a lower bound on $\langle h \rangle$ given by Eq.~\eqref{eq:phi+}, while $\phi_-$ an upper bound on $\langle h \rangle$ given by Eq.~\eqref{eq:phi-}. In terms of experimentally measured quantities, this means that $M_-/F_-\lesssim M_+/F_+ \simeq \theta_0 \lesssim \theta_{\rm exp} \simeq 10^{-10}$, where $\theta_0$ is the observed $\theta$-angle today, $\theta_{\rm exp}$ is given by the experimental upper bound on the neutron EDM~\cite{Abel:2020gbr}.  The electroweak scale is successfully selected if
\be
m_{\phi_+}^2 &\simeq& \frac{\Lambda_{\rm QCD}^4}{F_+^2}\, , \nn \\
m_{\phi_-}^2 &\simeq& \left(\theta + \frac{M_+}{F_+}\right)\frac{\Lambda_{\rm QCD}^4}{ F_- M_-} \, \gtrsim \, \frac{\Lambda_{\rm QCD}^4}{ F_-^2} \, , \label{eq:mphi}
\ee
where $\Lambda_{\rm QCD}^4\equiv \Lambda^4(v)\simeq (80~{\rm MeV})^4$. In this Letter we take $\mu_S \simeq \mu_B \simeq v$ to highlight the main features of the idea. In this limit $m_{\phi_\pm}$ are the physical masses of $\phi_\pm$ in our universe. If $m_{\phi_+} < \Lambda_{\rm QCD}^2/F_+$ in Eq.~\eqref{eq:potential} then $\mu_S < v$, but the physical mass of $\phi_+$ remains $\simeq \Lambda_{\rm QCD}^2/F_+$. For $\mu_B > v$ the physical mass of $\phi_-$ increases ($\Lambda_{\rm QCD} \to \Lambda(\mu_B)$ in Eq.~\eqref{eq:mphi}). This will be discussed more extensively in~\cite{DAgnoloTeresi}. In conclusion we have an ``axion" $\phi_+$ with a mass-coupling relation identical to the usual QCD axion. $\phi_+$ is solving the strong CP problem, but not in the usual way: all universes where $\theta > M_+/F_+$ crunch very quickly. The second scalar $\phi_-$ is an ALP heavier than a QCD axion with the same couplings. We comment more extensively on phenomenology after discussing in greater detail our cosmological dynamics.

Before discussing cosmology, let us point out that the flat direction of the QCD term in Eq.~\eqref{eq:phiG} does not pose a problem for the stability of the potential, independently of the Higgs vev. It is easy to show that for $M_-/F_- \lesssim \theta_0$ the flat direction of Eq.~\eqref{eq:phiG} is actually stabilized by the $\phi_-$ part of $V_\phi$.

\section{Cosmology}\label{sec:pheno}
We do not make any assumption on the Universe before reheating, except that it is one of many with different values of $\theta$, $m_h^2$ and the CC. We also do not make any assumption on the reheating temperature, which can range all the way from the BBN bound to $T_{\rm RH} \gtrsim M_{\rm GUT}$ without affecting our results. After reheating the cosmology of the model is determined by the temperature dependent values of $m_{\phi_\pm}$. For simplicity we imagine that $\phi_\pm$ are coupled to the reheating sector only through the QCD anomaly, so $V_\phi$ in Eqs.~\eqref{eq:potential} has its zero-temperature form throughout the history of the Universe, while $V_{\phi H}$ in Eq.~\eqref{eq:phiG} turns on at the QCD phase transition. 

\paragraph{Crunching Time} 
We can now solve
\be
\ddot \phi_\pm + 3 H \dot \phi_\pm +\frac{\partial V_\phi(\phi_\pm)}{\partial \phi_\pm}+\frac{\partial V_{\phi H}(\phi_\pm)}{\partial \phi_\pm}=0
\label{eq:dyn}
\ee
to determine the cosmological history of universes with different values of $m_h^2$ and $\theta$. We do not make any assumption on the initial position of $\phi_\pm$ along their potentials. We can first consider universes where initially $|\phi_\pm|\lesssim M_\pm$ and the Higgs vev or the $\theta$-angle are outside their stability interval Eq.~\eqref{eq:interval}. Their dynamics is determined entirely by the EFT in Eq.~\eqref{eq:potential}. It is easy to show that these universes do not enter a phase of $\phi_\pm$ inflation and $H$ is always dominated by radiation if the reheating temperature is compatible with bounds from BBN. 
Universes at the boundary, $\langle h \rangle \simeq \mu_B$ and $\langle h \rangle \simeq \mu_S$, give the longest crunching times in the multiverse. For these universes, we can solve Eq.~\eqref{eq:dyn} analytically. It takes $\phi_-$ a time $t_c^{\rm local} \simeq 1/m_{\phi_-}$ to cross the metastable region of size $\Delta \phi_- \simeq M_-$. Analogously, $\phi_+$ rolls down its potential on a time scale $t_c^{\rm local} \simeq 1/m_{\phi_+}$.
After crossing the local region, the slope of $V_\phi$ increases and it takes $\phi_\pm$ an additional time $t_c^{\rm global}$ to finally crunch. If the potential at $|\phi_\pm| \gtrsim M_\pm$ is dominated by the quartic term as in Eq.~\eqref{eq:potential}, a region $\Delta \phi$ is traversed in a short time $t_c^{\rm global} \sim \sqrt{\Delta \phi/V'} \sim \sqrt{M_\pm/\Delta \phi} / m_{\phi_\pm}$, after which the universe crunches. The kinetic energy accumulated during rolling redshifts away quickly $\sim 1/a^6$ and does not appreciably affect the crunching time~\cite{Strumia:2019kxg}. 

Similarly, universes that start from $|\phi_\pm|\gtrsim M_\pm$ always crunch, independently of $\langle h \rangle$ and $\theta$, in a short time $t_c^{\rm global}$. In conclusion, the longest crunching time in the multiverse is dominated by the local region $|\phi_\pm| \lesssim M_\pm$, and is of order $t_c \simeq \max[1/m_{\phi_+}, 1/m_{\phi_-}] \simeq 1/m_{\phi_+}$. 

To conclude we note that $m_{\phi_+}\lesssim H(\Lambda_{\rm QCD}) \simeq 10^{-11}$~eV~$\simeq (0.1~{\rm ms})^{-1}$, otherwise universes with the observed Higgs vev crunch before the stable minimum is generated at the QCD phase transition. However, while in this Letter we restrict to $\mu_S \sim v$, which determines the equality between $m_{\phi_+}$ in Eq.~\eqref{eq:potential} and Eq.~\eqref{eq:mphi} and the physical mass of $\phi_+$ in our universe, the lower bound on the Higgs vev could actually be much smaller, $\mu_S \ll v$. In this case, the crunching time is determined by the size of the potential in Eq.~\eqref{eq:potential} and can be longer than the inverse of the physical mass of $\phi_+$, set by $V_{\phi H}$. Then the region $m_{\phi_+} \gtrsim H(\Lambda_{\rm QCD})$ becomes allowed~\cite{DAgnoloTeresi}. 

In this picture our Universe is in a metastable, but extremely long-lived, state. The tunnelling action is very large $S \gtrsim \min_{\pm} [8 \pi^2 M_{\pm}^2/(3 m_{\phi_\pm}^2)]$
, giving a tunnelling time much longer than the age of the Universe.

\begin{figure}[!t]
$$
\includegraphics[height=0.4\textwidth]{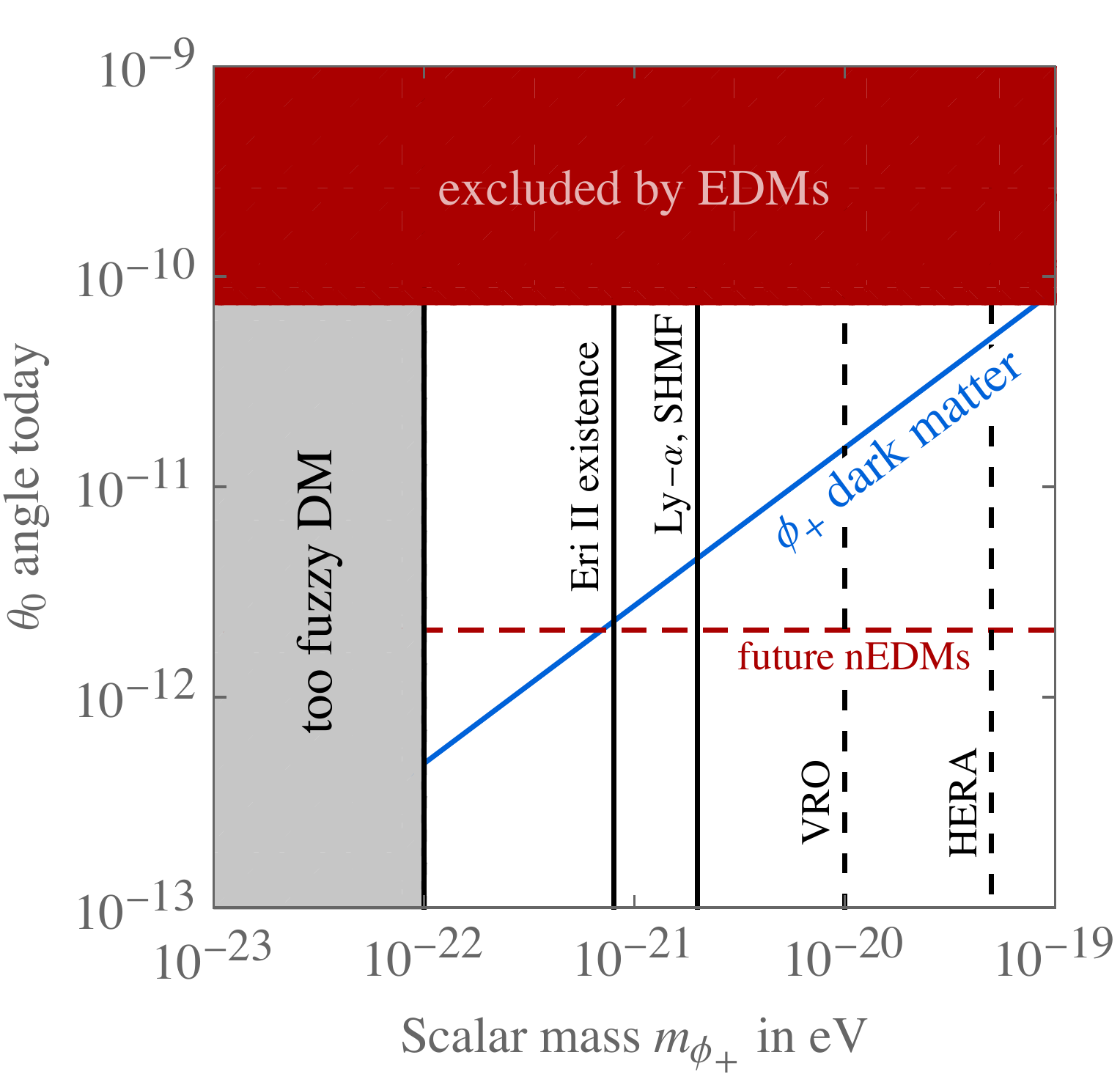}$$
\caption{Parameter space for  which $\phi_+$ constitutes the totality of DM of the Universe, as function of the $\phi_+$ mass and the $\theta$-angle today. Bounds~\cite{Abel:2020gbr} and future prospects~\cite{Abel:2018yeo,SNSnEDM,Filippone:2018vxf} from hadronic EDM searches are shown. New ideas involving molecular compounds could further improve the red dashed line~\cite{Hutzler:2020lmj, PhysRevA.100.032514, doi:10.1063/1.5098540, PhysRevC.94.025501, Yu:2020wtz}. We also plot constraints on fuzzy DM from 
Lyman-$\alpha$ forest~\cite{Leong:2018opi, Irsic:2017yje, Kobayashi:2017jcf, Armengaud:2017nkf, Bozek:2014uqa, Zhang:2017chj}, measurements of the subhalo mass function~\cite{Schutz:2020jox} and the Eridanus II dwarf galaxy~\cite{Marsh:2018zyw}. Tensions in the fit to other dwarf galaxies point to a similar constraint as Eridanus~\cite{Safarzadeh:2019sre, Bar:2018acw}. We shade the area where multiple observations disfavor the corresponding DM mass hypotheses~\cite{Hui:2016ltb}. The dashed lines correspond to the potential sensitivity from future observations in 21 cm cosmology (HERA)~\cite{Munoz:2019hjh} and by the Vera Rubin Observatory~\cite{Drlica-Wagner:2019xan}.\label{fig:parameter_space_phip}}
\end{figure}

\paragraph{Dark Matter} The scalars $\phi_\pm$ are stable over cosmological timescales ($\tau_{\phi_\pm}\simeq 10^{24}\;{\rm s}\; ({\rm eV}/m_{\phi_\pm})^5$) and their coherent oscillations can constitute the Dark Matter (DM) of the Universe~\cite{Preskill:1982cy,Abbott:1982af,Dine:1982ah}. To compute the relic density, we note that they get a ``kick" of order $\Delta \phi_{\pm} \simeq M_{\pm}$ at the QCD phase transition for universes where $\langle h \rangle\simeq \mu_{B,S}$. The maximum possible misalignment of $\phi_\pm$ from their safe minimum that does not entail a rapid crunch is $\mathcal{O}(M_\pm)$, so the relic density of $\phi_\pm$ is dominated by this kick. If our universe is close to one of the boundary values, modulo $\mathcal{O}(1)$ factors, we have $\rho_{\phi_\pm}(T\simeq \Lambda_{\rm QCD}) \simeq m_{\phi_\pm}^2 M_\pm^2$ for one of the two scalars. In this Letter we consider a scenario where $m_{\phi_-} \gg m_{\phi_+}$, and it is easy to show that $\phi_-$ is at most a subdominant component of DM. On the contrary $\phi_+$ can be the DM of our universe, with $\rho_{\phi_+}(T\simeq \Lambda_{\rm QCD}) \simeq m_{\phi_+}^2 M_+^2\simeq \theta_0^2 \Lambda_{\rm QCD}^4$. 
Since it starts oscillating and redshifting as cold DM when $H\simeq m_{\phi_+} < H(\Lambda_{\rm QCD})$, its relic density today is given by
\be
\frac{\rho_\phi}{\rho_{\rm DM}} 
\simeq \frac{\theta_0^2 \Lambda^4_{\rm QCD}}{T_{\rm eq} M_{\rm Pl}^{3/2} m_{\phi_+}^{3/2}}\simeq \left(\frac{\theta_0}{10^{-10}}\right)^2\left(\frac{10^{-19}\;{\rm eV}}{m_{\phi_+}}\right)^{3/2} \!\!\!\!\!\! , \label{eq:RD}
\ee
with $T_{\rm eq}\simeq {\rm eV}$ the temperature of matter-radiation equality. Therefore, its relic density is $\simeq \theta_0^2$ times smaller than the one of a Peccei-Quinn axion with the same mass, avoiding overclosure constraints of light axions. Taking $\mu_S \ll v$ leaves the relic density Eq.~\eqref{eq:RD} unaffected.

\paragraph{Inflation and Swampland Conjectures} Until now we have considered an anthropic selection mechanism for the CC, but we can also imagine that the CC is close to zero due to non-trivial constraints imposed in the IR by UV dynamics, as suggested by swampland conjectures. Differently from most of the existing approaches to cosmological naturalness, our mechanism is compatible with modern swampland conjectures~\cite{Palti:2019pca}. The form of the potential in Eq.~\eqref{eq:potential} is generic enough so that the de Sitter (dS) conjecture~\cite{Obied:2018sgi} is satisfied as long as $M_\pm < M_{\rm Pl}$, because in Planckian units the slope of the potential is not suppressed as compared to the potential itself. The observed small CC can be then simulated, as usual, by a cosmologically rolling scalar. The distance conjecture~\cite{Ooguri:2006in}  is satisfied if $M_\pm$ are sub-Planckian. Note that reproducing the observed dark matter relic density requires $F_+ > M_{\rm Pl}$, but the field excursions that we consider can be $\ll F_+$ and even for super-Planckian $F_+$  we might not violate the distance conjecture. Since this conjecture is argued to automatically imply the dS one~\cite{Garg:2018reu, Ooguri:2018wrx}, in this framework there are no large and positive CCs in the landscape and the global minimum of $\phi_\pm$ is guaranteed to crunch, without any stringent requirement on the splitting between the two minima.
In summary, the constraints imposed by these conjectures on the structure of the low-energy potential can be easily satisfied and those on the high energy landscape do not affect our results. 

However, we can consider, more generally, the relation between field excursions and the maximal CC for which our mechanism works. 
If we assume that the UV completion of Eq.~\eqref{eq:potential} does not affect its form until $\phi_\pm \sim M_{\rm Pl}$, then the maximal CC $\Lambda^4_{\rm sub}$ that does not require super-Planckian field excursions is 
\be
\Lambda_{\rm sub}^4 
\simeq \frac{\Lambda_{\rm QCD}^4}{\theta_0^2} \frac{M_{\rm Pl}^4}{F_+^4} \lesssim (10^{14} \GeV)^4 \left(\frac{10^{-10}}{\theta_0}\right)^2 \, ,
\ee
obtained as the potential at $\phi_+ \sim M_{\rm Pl}$, determined by the quartic coupling of $\phi_+$, using~\eqref{eq:mphi} and $M_+/F_+\simeq \theta_0$.
Here we took $\mu_S \ll v$, so that $F_+ \gtrsim 10^{8} \GeV$ from axion bounds~\cite{Zyla:2020zbs}, not restricted by $m_{\phi_+} \lesssim H(\Lambda_{\rm QCD})$.

We also find useful to stress that since our mechanism is compatible with standard inflation and does require a large number of $e$-folds, it does not implicitly reintroduce tuning or violation of swampland conjectures in other sectors of the theory. For the same reason it does not violate the considerations in~\cite{Dvali:2013eja, Dvali:2017eba} on the maximal time during which a semiclassical description of dS is valid. 

\section{Smoking-Gun Signals}\label{sec:signals}

The scalar $\phi_+$ is an axion of mass $m_{\phi_+} \lesssim 10^{-11} \eV$ which lies on the QCD line. Notice that $\phi_+$ does not give rise to black hole superradiance in the region $m_{\phi_+} \sim 10^{-12} \eV$ because of the self-coupling in Eq.~\eqref{eq:potential}~\cite{Baryakhtar:2020gao}. If observed in this region, this would then constitute a first characteristic trait that distinguishes $\phi_+$ from the Peccei-Quinn axion. 

If $m_{\phi_+} \lesssim 10^{-19} \eV$, $\phi_+$ can constitute the totality of  DM in our Universe, as shown in Fig.~\ref{fig:parameter_space_phip}. Its relic density is a function of the $\theta$-angle today and of $m_{\phi_+}$, as shown in Eq.~\eqref{eq:RD}. Limits on  fuzzy DM imply $\theta_0 \gtrsim 10^{-12}$, observable at future EDM experiments~\cite{Abel:2018yeo,SNSnEDM,Filippone:2018vxf}:  observing $\phi_+$ DM and small galaxies \emph{predicts} sizeable EDMs. Similarly, an observation of $\theta_0$ in the near future and a measurement of the DM mass would allow to test the smoking-gun relation in Eq.~\eqref{eq:RD}. A combination of future EDM measurements and fuzzy DM probes~\cite{Safarzadeh:2019sre, Bar:2018acw, Leong:2018opi, Irsic:2017yje, Kobayashi:2017jcf, Armengaud:2017nkf, Bozek:2014uqa, Zhang:2017chj, Hui:2016ltb,  Schutz:2020jox, Marsh:2018zyw,Munoz:2019hjh, Drlica-Wagner:2019xan} can fully test the hypothesis of $\phi_+$ DM, as shown in Fig.~\ref{fig:parameter_space_phip}.

We also predict a second axion-like particle with mass $m_{\phi_-} \gtrsim m_{\phi_+}$. $\phi_-$ is a ALP with mass comparable to or larger than a QCD axion with the same couplings, and, as such, is difficult to observe at current axion experiments, especially because it is not DM. However new experimental ideas could further corroborate this framework by finding a second ALP heavier than the one constituting DM.

To conclude, the two scalars $\phi_\pm$ have a very distinctive phenomenological structure and the pattern induced in axion and EDM experiments can provide a smoking-gun signal for the scenario that we propose. A partial or total future observation of this peculiar pattern would provide strong evidence towards the novel mechanism presented in this Letter that jointly addresses the Higgs hierarchy and strong-CP problems.

\mySections{Acknowledgments}
We thank M. Geller,  G. Giudice,  M. McCullough, R. Rattazzi, A. Strumia for discussions and comments on the manuscript.


\widetext
\bigskip\bigskip
\begin{center}
\LARGE \bf Appendix
\end{center}
\bigskip
\twocolumngrid

\section{Scalar Potential}
A concrete way to generate Eq.~\eqref{eq:potential} and UV complete it is as follows: $\phi_\pm$ can be part of an approximately scale-invariant sector in the UV. The Lagrangian is classically scale invariant, but the symmetry is broken by quantum effects as in QCD. Below some large scale $\Lambda_{\rm UV}$ the potential is in the form 
\be
V_\phi  =\lambda_\pm^\prime S_\pm^2 \phi_\pm^2+\lambda_\pm \phi_\pm^4\, , \label{eq:struc}
\ee
with $\lambda_\pm \sim \lambda^\prime_\pm \ll 1$. For concreteness here and in the following we take $M_\pm < F_\pm < \Lambda_{\rm UV}$, but other choices are compatible with the mechanism. In this case, at $F_\pm$ scale invariance is broken only by derivative couplings, such as $V_{\phi H}$, that preserve the shift symmetries $\phi_\pm\to \phi_\pm+c_\pm$ and do not generate other contributions to $V_\phi$. The final form of $V_\phi$ in Eq.~\eqref{eq:potential} is generated at $M_\pm$, by a spontaneous breaking of scale invariance by the vev of $S_\pm$, that gives $\phi_\pm$ a mass. Notice that possible additional couplings of the same order as $\lambda_\pm, \lambda^\prime_{\pm}$ (not written down explicitly in \eqref{eq:struc}) do not spoil the successful structure of the potential. For instance, a mixed quartic $\lambda^{\prime\prime}_\pm \phi_{\pm}^2 |H|^2$ gives a negligible contribution to the $\phi_\pm$ mass: $\lambda^{\prime\prime}_\pm v^2 \sim \lambda^{\prime}_\pm v^2 \ll  \lambda^{\prime}_\pm M_\pm^2 \equiv m_{\phi_\pm}^2$. Instead, mixed couplings $S_\pm^2 |H|^2$ just give additional contributions to the scanned Higgs mass in the landscape. The cross quartics $\phi_+^2 \phi_-^2$ generated by quantum corrections from Eq.~\eqref{eq:struc} do not change the picture in the main text, since they are negligibly small, $(\lambda^\prime_\pm)^2 \ll  \lambda^\prime_\pm$. The same is true for those generated by gravity. Even adding a term $\lambda_c \phi_+^2 \phi_-^2$ to the potential in Eq.~\eqref{eq:potential} would not appreciably change our conclusions provided that $\lambda_c \lesssim 2 \sqrt{\lambda_+ \lambda_-}$. So even cross couplings of the same order as couplings already present in our potential would not spoil the mechanism.
No further terms can be generated from the coupling to ${\rm Tr}[G\widetilde G]$, besides those considered in the main text. 

At the scale $\Lambda(\langle h \rangle)$ the shift symmetry is broken also by $V_{\phi H}$ down to $\phi_\pm \to \phi_\pm + 2\pi F_\pm n$ due to instanton effects. One might wonder if the non-compact nature of $V_\phi$ is posing a problem, given that $V_{\phi H}$ has this discrete gauge symmetry $\phi_\pm \to \phi_\pm + 2\pi F_\pm n$. First, notice that since we are in the regime $M_\pm \ll F_\pm$ this cannot pose an obstruction, because of decoupling arguments. More explicitly: the gauge symmetry can be either spontaneously broken in the UV or preserved by allowing for multiple branches of $V_\phi$~\cite{Dvali:2005an, Kaloper:2008fb} (e.g.~because of monodromy~\cite{Silverstein:2008sg, McAllister:2014mpa}). In this case Eq.~\eqref{eq:potential} implicitly assumes that $V_\phi$ is generated by CP-conserving dynamics in the UV. 

As a last ingredient, $V_\phi$ is stabilized in the UV either by higher-dimension operators at $\Lambda_{\rm UV}$, or by a small breaking of scale invariance, for instance in the form $\lambda^{\prime\prime\prime}_\pm \phi_\pm^{4+\epsilon}/M_\pm^\epsilon$, which generates a deep minimum at $\phi_\pm \sim M_\pm (\lambda^{\prime\prime\prime}_\pm/\lambda_\pm)^{1/\epsilon}$. The position of the minimum can naturally be $\gg M_\pm$.

It is useful to stress that the potential in Eq.~\eqref{eq:potential} and the UV completion that we have just described are just one of many possibilities that realize our idea. More general values of the $\mathcal{O}(1)$ factors in Eq.~\eqref{eq:potential}, as well as including tadpole and cubic terms (with scale $M_\pm$) and possibly cross-interactions, are compatible with the mechanism. Similarly scale invariance could be replaced by supersymmetry. We discuss more general implementations in a companion paper~\cite{DAgnoloTeresi}.

\section{Cosmological Constant Landscape}
The question of why the cosmological constant is finely scanned around zero in the metastable minimum of the $\phi_\pm$ potential is no different from the same question in any generic CC landscape. It does not require a coincidence of scales, but can be answered using symmetry.

In short, the minima of $\phi_\pm$ near the origin ($\phi_\pm=0$), which are affected by the Higgs, are generated by a small breaking of a symmetry, such as scale invariance or supersymmetry, as we discussed in the previous Section in the case of scale invariance. Therefore in these minima the CC is naturally small. A given landscape could scan this small CC around zero, but not the larger one in the other minima. 

To illustrate this point, consider the QFT toy model of a landscape in~\cite{ArkaniHamed:2005yv}. We imagine a theory with $N$ scalars $\phi_i$. Each scalar has a potential $V_{\phi_i}$ with two minima at $\langle \phi_i \rangle = \phi_{1,2}$ and vacuum energies $V_{1,2}$. We take $V_1 \geq V_2$. The full theory has $2^N$ vacua described by the potential
\be
V=\sum_{i=1}^N V_{\phi_i}\, . 
\ee
We can label the vacua using a set of integers $\eta_i=\pm 1$. Every choice of $\{\eta\}=\{\eta_1, ... , \eta_N\}$ corresponds to a different CC
\be
\Lambda_{\{\eta\}}&=& N \bar V + \sum_{i=1}^N \eta_i \Delta V\, , \nn \\ \bar V &=& \frac{V_1+V_2}{2}\, , \quad \Delta V=\frac{V_1-V_2}{2}\, .
\ee
For simplicity we have taken the same values of $V_{1,2}$ for all the scalars, since it does not affect our conclusions.

The distribution of CCs in the landscape at large $N$ is well approximated by a Gaussian (as expected from the central limit theorem)
\be
p(\Lambda)\to \frac{2^N}{\sqrt{2\pi N}\Delta V}e^{-\frac{\left(\Lambda - N \bar V\right)^2}{N\Delta V^2}}\, .
\ee
If we have enough minima to populate only the central region of the Gaussian, the CC is finely scanned in a region $\Lambda = \bar \Lambda \pm \delta \Lambda = N \bar V \pm \sqrt{N} \Delta V$. If $\bar V \simeq \Delta V$, as we expect from dimensional analysis, then
\be
\frac{\delta \Lambda}{\bar \Lambda} \simeq \frac{1}{\sqrt{N}}\, .
\ee
In particular we are not scanning around zero in the central region of the Gaussian. In this landscape the number of vacua with nearly vanishing vacuum energy is $\simeq 2^N e^{- N \bar V^2/\Delta V^2}$. To finely scan the CC around zero we need both $\bar V/\Delta V \lesssim \sqrt{\log 2}$ and sufficiently large $N$. A generic landscape is finely scanning the CC only around $N \bar V$.

The situation is different in supersymmetry. Take for instance the odd superpotential
\be
W=\lambda \phi^3 -\mu^2 \phi\, .
\ee
In this case at the two minima $W_1 = - W_2$ so that $\bar W=0$. Then the landscape generated by $N$ of these superpotentials is scanning the CC in the range
\be
- 3 \frac{|\sqrt{N} \Delta W|^2}{M_{\rm Pl}^2} \lesssim \Lambda \lesssim 0 \, .
\ee
In this case supersymmetry is keeping $\Lambda\leq 0$ and a $Z_4$ R-symmetry that protects the odd structure of the superpotential is ensuring that the distribution of negative CCs has a central value comparable to its standard deviation: $|\sqrt{N} \Delta W|^2/M_{\rm Pl}^2$~\cite{ArkaniHamed:2005yv}. After SUSY breaking, this landscape scans the CC efficiently around zero, because of its symmetries.

We can now add our two scalars $\phi_\pm$ to the landscape: 
\be
V_\phi=V_+(\phi_+)+V_-(\phi_-)\, . 
\ee
As in stable universes in the main body of the paper, $V_\pm(\phi_\pm)$ have each a minimum near the origin (that we set to $\phi_\pm=0$ without loss of generality) and a second minimum at large values of $\phi_\pm^{\rm min}$, where $V_\pm(\phi_\pm^{\rm min})<0$. We indicate with $\Delta V_\pm = V_\pm(0)-V_\pm(\phi_\pm^{\rm min}) >0$ the difference in vacuum energy between the minima.

Obviously the non-supersymmetric case is unaltered. We have just added two more scalars to the landscape and we do not expect to scan the CC around zero in the absence of accidental tunings.

Now consider the supersymmetric case and let us take $\Delta V_\pm$ much larger than the typical width of the CC distribution in the landscape, as we did in the main body of the paper, i.e. $\Delta V_\pm \gg |\sqrt{N} \Delta W|^2/M_{\rm Pl}^2$. The only natural explanation of a small CC is that $|V_{\pm}(0)|$ are small because of a symmetry. In particular if $|V_{\pm}(0)| \lesssim |\sqrt{N} \Delta W|^2/M_{\rm Pl}^2 \ll \Delta V_\pm$ the CC can be scanned around zero, but only in the minimum of $V_\phi$ near the origin, whose properties are affected by the Higgs. 

We have already introduced an example of the required symmetry in the previous Section: the $\phi_\pm$ sector could be approximately scale invariant below the scale of its deep minimum. In this case $V_{\pm}(0)$ are naturally of the order of the breaking of scale invariance $|V_{\pm}(0)|\sim M_\pm^4 \ll \Delta V_\pm$.

Further note that no coincidence between $M_\pm^4$ and $|\sqrt{N} \Delta W|^2/M_{\rm Pl}^2$ is required, in particular the CC is scanned around zero even if $M_\pm^4 \ll |\sqrt{N} \Delta W|^2/M_{\rm Pl}^2$. 

The same results can be obtained in a number of different ways. For instance in~\cite{DAgnoloTeresi} we discuss a case without scale invariance where $\phi_\pm$ are approximately supersymmetric.

In summary, to naturally scan the CC around zero we always need a symmetry: in our framework this symmetry is approximately conserved in one minimum of $V_\pm$, but badly broken in the other one.


\providecommand{\href}[2]{#2}\begingroup\raggedright
\endgroup

\end{document}